\documentclass[hyper,letterpaper]{JHEP3} 

\usepackage{epsfig,multicol,bbm}

\newcommand{\reef}[1]{(\ref{#1})}
\DeclareMathSymbol{\IR}{\mathbin}{AMSb}{"52}

\title{Global R--Currents and Phase Transitions in Large $N_c$ Gauge Theory}

\author{{Tameem Albash, Veselin Filev, Clifford V. Johnson, Arnab Kundu}\\
	Department of Physics and Astronomy\\ University of Southern California\\ Los Angeles, CA 90089\\
	E-mail: \email{albash@usc.edu}, \email{filev@usc.edu}, \email{johnson1@usc.edu}, \email{akundu@usc.edu}}

\preprint{\hepth{0605175}}	

\abstract{We study, using a gravity dual, the finite temperature dynamics of $SU(N_c)$ gauge theory for large $N_c$, with fundamental quark flavours in a quenched approximation, in the presence of a fixed R--charge under a global R--current.  We observe several notable phenomena. There is a first order phase transition where the quark condensate jumps discontinuously at finite quark mass, generalizing similar transitions seen at zero charge.  Our tool in these studies is holography, the string dual of the gauge theory being the geometry of $N_c$ spinning D3--branes at finite temperature, probed by a D7--brane.}


\begin{document} 


\section{Introduction}
There is great interest in understanding the phase structure of Quantum Chromodynamics (QCD). There are various regions of the $(\mu_B,T)$ phase diagram ($T$ is temperature and $\mu_B$ is a chemical potential for baryon number) which are extremely hard to model theoretically using traditional field theory techniques. Some of these regimes are being probed experimentally, and so there is additional interest in obtaining better theoretical understanding from an immediate phenomenological point of view.

For almost a decade, progress in understanding the dynamics of extended objects in string theory has enabled several new techniques--and a fresh and powerful new perspective--to be brought to the study of strongly coupled gauge theory. The techniques were sharpened further (for large number of colours, $N_c$,) with holographic examples such as the AdS/CFT correspondence, and the study of gauge/gravity duals (or, more properly, gauge/string duals) has proven to be a fruitful and often elegant pursuit.

Not long after the birth of AdS/CFT~\cite{Maldacena:1997re,Witten:1998qj,Gubser:1998bc}, where the finite temperature phase structure of ${\cal N}=4$ $SU(N_c)$ (at large $N_c$) was understood in terms of the thermodynamics of Schwarzschild black holes in AdS~\cite{Witten:1998qj,Witten:1998zw}, it was recognized~\cite{Chamblin:1999tk} that progress could be made in understading the properties of gauge theories in the presence of a global current, akin to a chemical potential for baryon number or isospin, by studying the physics of charged black holes, such as Riessner--Nordstrom~\cite{Chamblin:1999tk,Chamblin:1999hg}, or more general charged black holes with non--trivial scalars~\cite{Cvetic:1999rb}, in AdS.

The resulting $(\mu, T)$ or $(q,T)$ phase diagrams (where $q$ is conjugate to $\mu$--the analogue of Baryon number) of the ${\cal N}=4$ guage theory were found to be rather rich, with {\it e.g.,} a first order phase transition line, and even a second order critical point in the $(q,T)$ plane~\cite{Chamblin:1999tk,Chamblin:1999hg}. This led to the hope\footnote{C. V. Johnson, various seminars, 2000.} that such studies might lead to insights into the physics of the QCD phase diagram. The idea is that studies of the effect of the global $U(1)$ symmetry--even though it is not exactly baryon number--might lead to physics in the same universality class as the more realistic gauge theories, giving insight into QCD at finite temperature and density.

A more firm footing for this idea should be obtained by the study of those dynamics in the presence of fundamental flavours of quark, a key feature of QCD that distinguishes it from the ${\cal N}=4$ gauge theory (which of course only has adjoint matter of very specific types and quantities allowed by supersymmetry). While it is notoriously difficult to study this problem as a string dual ---so far--- it is known~\cite{Karch:2002sh} how to place fundamental flavours into the theory, in the limit $N_c{\gg}N_f$. This is a sort of ``quenched'' limit where the quarks feel the gauge dynamics but whose presence does not effect the physics ({\it i.e.,} the quarks do not back--react).

While this limit cannot capture a great deal of detailed QCD physics,
it is hoped that it may well provide valuable clues in the study of
the phase structure and will certainly be an instructive step along
the path to fully understanding the case of fully interacting
dynamical quarks.

The global symmetries in question in the case of the ${\cal N}=4$ theory live in the $U(1)^3$ subgroup of the large $SO(6)\subset SU(4)$ R--symmetry enjoyed by the theory. In the string dual this corresponds to that fact that the D3--branes (on whose worldvolume the gauge theory resides) have an $\IR^6$ transverse to them, In the decoupling limit that gives the AdS/CFT duality, they correspond to a subset of the $SO(6)$ isometries of the explicit $S^5$ of the AdS$_5\times S^5$ spacetime where the dual string theory propagates.

Fundamental dynamical quarks enter the story as D7--brane probes of the geometry. The ${\cal N}=4$ supersymmetry is now replaced by ${\cal N}=2$ and the quarks lie in hypermultiplets. The $\IR^6$ now breaks into an $\IR^2$ transverse to both the D3--branes and the D7--branes, and an $\IR^4$ transverse to the D3--branes, but internal to the D7--branes.

In the probe limit in question, the quarks are the ends of the strings stretching from the D7--branes to the D3--branes, their masses being set by the distance between the two sets of branes in the $\IR^2$. In the decoupling limit (where the $N_c$ D3--branes have been replaced by AdS$_5\times S^5$ geometry), the location in the plane transverse to the D7--brane controls the quark mass and a condensate of quarks, according to~\cite{Kruczenski:2003uq}:
\begin{equation} \label{eqt: L}
\lim_{u \rightarrow \infty} L(u) \sim m + \frac{c}{u^2} +\cdots\ ,
\end{equation}
where $m = 2 \pi \alpha' m_q$ (the bare quark mass) and $-c = \langle \bar{\Psi} \Psi \rangle / (8 \pi^3\alpha' N_f \tau_{\mathrm{D7}})$ (the quark condensate), $u$ is a radial coordinate in the decoupling limit, and $N_f = 1$ in this paper.

The condensate actually breaks the $U(1)$ symmetry corresponding to rotations in the $\IR^2$, that acts chirally on the $\Psi$ fields. We explored the allowed values of the condensate as a function of quark mass $m$ at finite temperature $T$ (where supersymmetry is broken), and with the system charged an amount $q$ under a global $U(1)$ current. We focused on two cases. The first is the chiral current associated with rotations transverse to the D7--branes, and the second is a $U(1)$ that mixes transverse and parallel rotations.

We found a rich structure very similar to the case studied in ref.~\cite{Albash:2007bq}, upon which we will report in this paper\footnote{Note that this is a thoroughly revised version of an earlier manuscript~ \cite{Albash:2006bs}. Several key physical results have changed as a result of our improved understanding of several issues, following work in refs.~\cite{Karch:2007pd,Albash:2007bq}.}.  The induced geometry on the D7--brane has an effective horizon that lies above the event horizon of the geometry; for the physical questions we ask, the physics beneath this effective horizon is screened from an observer at the boundary.
In order to satisfy Ramond--Ramond charge conservation of the D7--brane, extending the D7--brane across the effective horizon complexifies the mass term in the boundary gauge theory.
We found that there are generalizations of the first order phase transition already observed (and extensively studied, see refs.~\cite{Albash:2006ew, Babington:2003vm, Karch:2006bv, Kirsch:2004km, Ghoroku:2005tf, Mateos:2006nu}) at $q=0$ for this case.  


\section{Spinning D3-branes}
\label{sec:spinning D3 branes}
To source subgroups of the $U(1)^3\subset SO(6)$ global symmetries corresponding to rotations in the transverse $\IR^6$, we consider D3--branes with angular momentum.  The supergravity description/metric for spinning D3--branes is~\cite{Kraus:1998hv,Cvetic:1999xp}:
\begin{eqnarray}
ds^2 &=& H_3^{-1/2}\left\{-\left(1-\frac{r_H^4}{r^4 \Delta}\right)dt^2 + d\vec{x}\cdot d\vec{x}\right\} +H_3^{1/2} \left\{\frac{\Delta dr^2}{\mathcal{H}_1 \mathcal{H}_2 \mathcal{H}_3 -\frac{r_H^4}{r^4}} +r^2\sum_{i=1}^3\mathcal{H}_i \left(d\mu_i^2 +\mu_i^2 d\phi_i^2\right) \right.\nonumber\\
& & \left.-\frac{2r_H^4
\cosh\left(\beta_4\right)}{r^4 H_3 \Delta}dt \left(\sum_{i=1}^3
R_i\mu_i^2 d\phi_i\right) +\frac{r_H^4}{r^4}\frac{1}{H_3
\Delta}\left(\sum_{i=1}^3 R_i \mu_i^2 d\phi_i\right)^2 \right\} \ ,
\label{spinninng D-branes}
\end{eqnarray}
where:
\begin{eqnarray}
\Delta &=& \mathcal{H}_1 \mathcal{H}_2 \mathcal{H}_3 \sum_{i=1}^3
\frac{\mu_i^2}{\mathcal{H}_i} \nonumber\\
H_3 &=& 1+\frac{r_H^4}{r^4}\frac{\sinh\left(\beta_3\right)}{\Delta}
= 1+\frac{\alpha_3 r_3^4}{\Delta r^4} \nonumber\\
\mathcal{H}_i &=& 1+\frac{\ell_i^2}{r^2} \nonumber\\
r_3^4&=&4 \pi g_s N \alpha'^2 = \alpha'^2 \hat{r}^4 \nonumber\\
\alpha_3 &=&\sqrt{1+\left(\frac{r_H}{2r_3}\right)^8}
-\left(\frac{r_H}{2r_3}\right)^4 \nonumber \ .
\end{eqnarray}
In order to explore the decoupling limit ($\alpha' \to 0$), which we are interested in, we define:
\begin{eqnarray}
r &=& \alpha ' u\ ,\quad 
r_H = \alpha' u_H,\quad 
\ell_i =\alpha' q_i \nonumber \ ,
\end{eqnarray}
such that:
\begin{eqnarray}
\frac{\alpha_3 r_3^4}{r_H^4} &\to& \frac{\hat{r}_3^4}{\alpha'^2
u_H^2}\ ,\quad 
\sinh\left(\beta_3\right) \to \frac{\hat{r}_3^2}{\alpha' u_H^2}\ ,
\nonumber\\
\mathcal{H}_i &\to& 1+\frac{q_i^2}{u^2}\ , \quad 
H_3 \to \frac{\hat{r}_3^4}{\alpha'^2 u^4 \Delta} \nonumber \ .
\end{eqnarray}
With these definitions, in the decoupling limit, the metric is:
\begin{eqnarray} 
\label{eqt: decoupled spinning brane metric}
ds^2/\alpha' &=& \Delta^{\frac{1}{2}}\left\{-\left(\mathcal{H}_1
\mathcal{H}_2 \mathcal{H}_3\right)^{-1} f dt^2 + f^{-1} du^2 +
\frac{u^2}{R^2}d\vec{x}\cdot d\vec{x}\right\} \nonumber\\
&&+\Delta^{-\frac{1}{2}}\left\{\sum_{i=1}^3\mathcal{H}_i \mu_i^2
\left(R d\phi_i - A_t^i dt\right)^2 + R^2 \mathcal{H}_i d\mu_i^2\right\} \ ,
\end{eqnarray}
where we have defined:
\begin{eqnarray}
R &=& \hat{r}_3\ ,\quad 
f=\frac{u^2}{R^2}\mathcal{H}_1\mathcal{H}_2\mathcal{H}_3-
\frac{u_H^4}{u^2 R^2}\ , \quad 
\mbox{\rm and }\quad A_t^i = \frac{u_H^2}{R \mathcal{H}_i}\frac{q_i}{u^2} \ .
\end{eqnarray}
The event horizon, which we denote by $u_E$, is given by the largest root of $f(u) = 0$.  In terms of $z= u^2$, $f(z)$ is a cubic equation, which motivates us to write $f(u)$ in terms of the three roots of $f(z)$:
\begin{equation}
f(u) = \frac{1}{R^2 u^4} \left(u^2 - u_E^2 \right) \left(u^2-u_1^2 \right) \left(u^2 -u_2^2 \right) \ ,
\end{equation}
where $u_1^2$ and $u_2^2$ are the other roots of $f(z)$ and may not be real.  In this notation, the temperature of the background (and the dual gauge theory) is given by~\cite{Russo:1998by}:
\begin{equation} \label{eqt:temperature}
T = \frac{u_E}{2 \pi R^2 u_H^2} \left(u_E^2 - u_1^2 \right)\left(u_E^2 - u_2^2 \right) \ .
\end{equation}
The geometry as given by the metric in equation \reef{eqt: decoupled spinning brane metric} has an ergosphere given by:
\begin{equation}
 - \Delta \left(\mathcal{H}_1 \mathcal{H}_2 \mathcal{H}_3 \right)^{-1} f + \sum_{i=1}^3 \mathcal{H}_i \mu_i^2 \left(A_t^i \right)^2 = 0
\end{equation}
The ergosphere can be ``removed'' by changing coordinates to a rotating frame, which is equivalent to gauge shifting $A_i$:
\begin{equation}
\phi_i ' = \phi_i - \mu_R^i t \quad \leftrightarrow \quad A_t^{\prime i} = A_t^i - R \mu_R^i \ ,
\end{equation}
where $\mu_R^i$ is set by requiring $A_t^{\prime i}$ to be zero at the event horizon:
\begin{equation} \label{eqt:mu_R}
\mu_R^i = \frac{u_H^2}{R^2} \frac{q_i}{u_E^2 + q_i^2} \ .
\end{equation}
In these coordinates, the geometry is no longer simply asymptotically AdS$_5 \times S^5$; it includes rotations along the three circles given by $\phi_i$ with angular velocity $\mu_R^i$ respectively.  In the dual gauge theory, this corresponds to having a time--dependent phase for the adjoint complex scalars; this is equivalent to having an R--charge chemical potential for the same scalars~\cite{Chamblin:1999tk}.
\section{Introducing Fundamental Matter}
We now proceed to introduce fundamental matter into the dual gauge theory.  This is done by introducing $N_f$ parallel D7--branes into the supergravity geometry~\cite{Karch:2002sh}, which introduces $\mathcal{N}=2$ hypermultiplets into the dual gauge theory.  In particular, the D7--branes are taken to be in the probe limit, i.e. where $N_f \ll N_c$ and the D7--branes do not back react on the geometry, which corresponds to the ``quenched'' approximation in the dual gauge theory.  We will work with $N_f =1$.

A complication arising from the geometry is that there is no undeformed $S^3$ for the D7--branes to wrap, unless $q_2 = q_3$.  This choice restores part of the symmetry of the geometry by having an $S^3$---parametrized by $\psi, \ \phi_2, \  \rm{ and }  \ \phi_3$---inside the deformed $S^5$.  With this less general background in mind, we introduce the D7--brane probe along the AdS and $S^3$ directions.  The brane is free to move in the $\theta$ and $\phi_1$ direction.  In particular, if the D7--brane is allowed to co--rotate with the same angular velocity as the background along the $\phi_1$ direction, this should be interpreted as having introduced the same R--charge chemical potential for the fundamental and the adjoint matter.

There are two different cases that we consider:
\begin{enumerate}
\item A single charge in the $\phi_1$ direction: $q_1 = q$ and $q_2 = q_3 =0$.  This case is a useful test case since it is the simplest problem to explore and provides valuable insight into the next, more complicated case.
\item Three equal charges: $q_1 = q_2 = q_3 = q$.  This case is perhaps of greater interest since the supergravity metric, upon compactification, corresponds to the AdS--Reissner--Nordstr\"om black hole~\cite{Chamblin:1999tk} (the previous case gives charged black holes with an extra scalar field excited~\cite{Cvetic:1999rb}).
\end{enumerate}
\section{Single Charge}
\label{sec: single charge}
\subsection{Properties of the embeddings}
For the single charge case, we choose $q_1 = q$ and $q_2 = q_3 =0$ in equation~\reef{eqt: decoupled spinning brane metric}, and the background metric (without the ergosphere) becomes:
\begin{eqnarray} \label{eqt: single charge metric}
ds^2 / \alpha'&=&\Delta^{1/2} \left\{ -\mathcal{H}^{-1} f dt^2 + \frac{u^2}{R^2} d\vec{x}\cdot
d\vec{x}\right\} +\Delta^{1/2} \left\{f^{-1} du^2 +R^2 d\theta^2\right\} \nonumber \\
&&+ \Delta^{-1/2}\left\{\mathcal{H} \sin^2\theta\left(R d\phi'_1 - A_t^{\prime 1} dt\right)^2 + R^2\cos^2\theta d\Omega_3^2\right\} \ ,
\end{eqnarray}
where
\begin{equation}
f = \frac{u^2}{R^2}\mathcal{H}-\frac{u_H^4}{u^2 R^2} \ , \quad 
\mathcal{H} = 1+ \frac{q^2}{u^2} \ , \quad 
A_t^{\prime 1} = \frac{u_H^2 q}{R \left(u^2 +q^2 \right)} -  \frac{u_H^2 q}{R \left(u_E^2 +q^2 \right)}  \ , \quad
\Delta = 1+ \frac{q^2}{u^2} \cos^2\theta \ . \nonumber
\end{equation}
The event horizon $u_E$ satisfies the equation:
\begin{equation}
u_E^2 \left(u_E^2 + q^2 \right) - u_H^4 = 0
\end{equation}
To proceed, we consider the action for the probe D7--brane to second order in $\alpha'$:
\begin{eqnarray}
\frac{S_{\mathrm{D7}}}{N_f}&=& S_{\mathrm{DBI}} + S_{\mathrm{WZ}} \nonumber \\
&=& - T_{\mathrm{D7}} \int d^8 \xi \ \mathrm{det}^{1/2} \left(P\left[ G_{a b} \right] + P \left[B_{a b} \right] + 2 \pi \alpha' F_{a b} \right) \nonumber\\ &&\hskip2cm + \left(2 \pi \alpha' \right)^2  \frac{\mu_7}{2} \int F_{(2)} \wedge F_{(2)} \wedge P\left[ C_{(4)} \right] \ ,\nonumber
\end{eqnarray}
where $P\left[ G_{a b} \right]$, $P\left[ B_{a b} \right]$, and $P\left[ C_{(4)} \right]$ are the pullback of the background metric, the $B$--field, and the 4--form potential (sourced by the $N$ D3--branes) onto the D7--brane worldvolume respectively, $F$ is the field strength on the D7--brane worldvolume, and $ T_\mathrm{D7} =\mu_7 / g_s $.  We begin by considering an ansatz of the form:
\begin{equation} \label{eqt:ansatz}
\theta = \theta(u) \ , \quad \phi'_1 = \phi_1(u) - \frac{u_H^2 q}{R \left(u_E^2 +q^2 \right)} t \ .
\end{equation}
where we have allowed the D7--brane to extend and co--rotate with the background along the circle $\phi'_1$.  The action is given by:
\begin{equation} \label{eqt:action}
S_{\mathrm{D7}}=- 2 \pi^2 u_H^4 N_f T_{\rm D7}  \int d^4 x \int d \tilde{u} \ \tilde{\mathcal{L}} \ ,
\end{equation}
where we have defined dimensionless parameters 
\begin{equation} \label{eqt:dimenionless}
u = u_H \tilde{u} \ , \quad q = u_H \tilde{q} \ , \quad \theta(u) = \theta(\tilde{u}) \ , \quad \phi_1(u) = \phi_1(\tilde{u}) \ , \quad \tilde{f}(\tilde{u}) = \tilde{u}^4 + \tilde{q}^2 \tilde{u}^2- 1 \ ,
\end{equation}
and
\begin{equation}
\tilde{\mathcal{L}} =  \tilde{u}^3 \cos^3 \theta(\tilde{u}) \sqrt{\left( 1 - \frac{\tilde{u}^2 \tilde{q}^2}{ \tilde{f}(\tilde{u})} \sin^2 \theta(\tilde{u}) \right) \left(1+ \frac{\tilde{f}(\tilde{u})}{\tilde{u}^2} \theta'(\tilde{u})^2 \right) + \frac{\tilde{f}(\tilde{u})}{\tilde{u}^2} \sin^2 \theta(\tilde{u}) \phi_1'(\tilde{u})^2} \ ,
\end{equation}
The equation of motion for $\phi_1$ can be integrated to give us a constant of motion:
\begin{equation}
\frac{\tilde{u} \cos ^3 \theta \tilde{f}(\tilde{u}) \sin^2 \theta \phi_1'(\tilde{u})}{\sqrt{\left( 1 - \frac{\tilde{q}^2 \tilde{u}^2}{ \tilde{f}(\tilde{u})} \sin^2 \theta(\tilde{u}) \right) \left(1+ \frac{\tilde{f}(\tilde{u})}{\tilde{u}^2} \theta'(\tilde{u})^2 \right) + \frac{\tilde{f}(\tilde{u})}{\tilde{u}^2} \sin^2 \theta(\tilde{u}) \phi_1'(\tilde{u})^2}}= - \tilde{K} \ .
\end{equation}
This expression can be inverted to give an expression for $\phi_1'$:
\begin{equation}
\phi_1'(\tilde{u}) = - \frac{\tilde{K} }{\sin \theta(\tilde{u}) \tilde{f}(\tilde{u})} \sqrt{\frac{\left(\tilde{f}(\tilde{u}) - \tilde{q}^2 \tilde{u}^2 \sin^2 \theta(\tilde{u}) \right) \left(\tilde{u}^2+ \tilde{f}(\tilde{u}) \theta'(\tilde{u})^2 \right)}{\left(-\tilde{K}^2 + \tilde{u}^4 \tilde{f}(\tilde{u}) \cos^6 \theta(\tilde{u}) \sin^2 \theta(\tilde{u}) \right)} } \ .
\end{equation}
With this result, we can choose to work with $\tilde{K}$ as a variable instead of $\phi_1$ by performing a Legendre transformation:
\begin{eqnarray} \label{eqt:legendre}
\tilde{I} &=& \tilde{\mathcal{L}} - \frac{\partial \tilde{\mathcal{L}}}{\partial \phi_1'} \phi_1' \\
&=& \frac{1}{\tilde{f}(\tilde{u}) \sin \theta(\tilde{u})} \sqrt{\left(-\tilde{K}^2 + \tilde{u}^4 \tilde{f}(\tilde{u}) \cos^6 \theta(\tilde{u}) \sin^2 \theta(\tilde{u}) \right) \left(\tilde{f}(\tilde{u}) - \tilde{q}^2 \tilde{u}^2 \sin^2 \theta(\tilde{u}) \right) \left(\tilde{u}^2+ \tilde{f}(\tilde{u}) \theta'(\tilde{u})^2 \right)} \nonumber 
\end{eqnarray}
In order for this object to remain real, the first two terms under the square root must either be both positive or both negative, meaning that they must change sign at the same time.   This requirement fixes a special radius in the geometry, which we denote by $\tilde{u}_\ast$, given by the largest real root of the equation:
\begin{equation} \label{eqt:vanishing_locus}
\tilde{u}^4 +\tilde{q}^2 \tilde{u}^2 \cos^2 \theta(\tilde{u}) - 1 = 0 \ .
\end{equation}
This in turn fixes the value of $\tilde{K}$ to be:
\begin{equation} \label{eqt:k1}
\tilde{K}^2 = \tilde{u}_\ast^4 \tilde{f}(\tilde{u}_\ast) \cos^6 \theta_0 \sin^2 \theta_0 \ ,
\end{equation}
where we have defined $\theta_0 \equiv \theta (\tilde{u}_\ast)$.  We leave the explanation of the physical significance of $\tilde{K}$ for later.  We refrain from writing out the equation of motion derived from the Legendre transformed Lagrangian in equation \reef{eqt:legendre}, but we write the asymptotic behavior of $\theta$:
\begin{equation} \label{eqt:asymptotic}
\lim_{\tilde{u} \to \infty} \theta(\tilde{u}) = \frac{1}{\tilde{u}} \left( \tilde{m} + \frac{\tilde{c}}{\tilde{u}^2} + O\left(\frac{1}{\tilde{u}^3} \right) \right) \ .
\end{equation}
From the AdS/CFT dictionary, this behavior corresponds to a dimension 3 operator with a source term proportional to $\tilde{m}$ and with a vev proportional to $\tilde{c}$.  The source is given by the bare quark mass and is given by:
\begin{equation} \label{eqt:m}
m_q = \frac{u_H \tilde{m}}{2 \pi \alpha'} \ ,
\end{equation}
and the vev of the operator, which we refer to as the condensate, is given by:
\begin{equation} \label{eqt:condensate}
\langle \bar{\Psi} \Psi \rangle = - 8 \pi^3 \alpha' V N_f T_{\rm D7} u_H^3 \tilde{c}'  \ ,
\end{equation}
where $\tilde{c}' = \tilde{c} + \tilde{m} \frac{\tilde{q}^2}{2}$.  An explanation of the additional term in $\tilde{c}'$ is given in Appendix \ref{appendix:condensate}.
%
%

We find that the solutions separate into two classes of solutions.  For $\theta_0 < \pi/2$, the embeddings cross the vanishing locus and fall into the event horizon.  These embeddings are referred to as ``black hole'' embeddings, and we show later that they correspond to ``dissociated'' mesons.  As was found in ref.~\cite{Albash:2007bq}, there exists a minimum $\theta_0$ after which the black hole embeddings, when they cross the vanishing locus, develop a conical singularity before striking the event horizon.   It is possible that string theory corrections may resolve the singularity, but this has not been shown yet.  However, as was argued in ref.~\cite{Albash:2007bq}, the phase diagram is not affected by the presence of these embeddings.  For $\theta(\tilde{u}_{\rm min}) = \pi /2$ with $\tilde{u}_{\rm min} > \tilde{u}_\ast$, the embeddings close, meaning that the $S^3$ that the D7--brane wraps shrinks to zero size,  before reaching the event horizon.  These embeddings are referred to as ``Minkowski'' embeddings, and these correspond to stable mesons.  The two types of solutions are separated by a critical embedding given by $\theta_0 = \pi/2$.  We present an example of the three possible situations in figure \ref{fig:1q_embeddings}.
\FIGURE{\epsfig{file=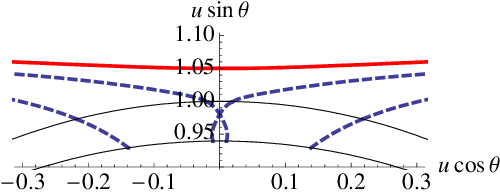,width=9cm}
 \caption{\small The dashed blue curves are black hole embeddings that cross the vanishing locus (denoted by the upper solid, thin black curve) and hit the event horizon (denoted by the lower solid, thin black curve).  One of the black hole embeddings crosses the $y$--axis, which means it develops a conical singularity.  The solid red curve is a Minkowski embedding.}
 \label{fig:1q_embeddings}}
\subsection{Analytical Results: Large $\tilde{q}$ limit}
We focus on Minkowski embeddings in the case where $\tilde{q} \gg 1$.  Returning to dimensionful coordinates, this limit corresponds to $q \gg u_H$.  In this limit, given equation \reef{eqt:dimenionless}, we can consider taking~$\tilde{f} \approx \tilde{u}^4 + \tilde{q}^2 \tilde{u}^2$ and $A_t'^{1} \approx 0$.  It is convenient to change coordinates to $\tilde{\rho} = \tilde{u} \cos \theta$ and $\tilde{L} = \sqrt{\tilde{u}^2 + \tilde{q}^2} \sin \theta$, which when taking with the ansatz $L \equiv L(\rho)$, gives us the exact Lagrangian for the pure AdS$_5 \times S^5$ case:
\begin{equation}
\mathcal{L} \propto \rho^3 \sqrt{1+ L'(\rho)} \ .
\end{equation}
This has the known solutions~\cite{Kruczenski:2003be} $L(\rho) = \sqrt{u^2 + q^2} \sin \theta = \rm constant$.  The constant is determined by requiring the D7--brane to close at some $u_{\rm min}$ with $\theta( u_{\rm min}) = \pi/2$.  Therefore, in the limit of large $q$, we can relate the bare quark mass to $u_{\rm min}$ via:
\begin{equation}
m = \sqrt{u_{\rm min}^2 + q^2}  \ .
\end{equation}
We can use this result to estimate the mass at which the phase transition occurs by finding the mass of the critical embedding (the borderline case between Minkowski and black hole embeddings), which has $u_{\rm min} = u_H$.  While it is true that the critical embedding is by--passed by the first order phase transition, we expect that the difference between the two is not too great, and find this to be the case numerically (see later).  Therefore, for the critical embedding, we have (in dimensionless variables):
\begin{equation} \label{eqt:m_star}
\tilde{m}_\ast = \sqrt{1+ \tilde{q}^2} \ .
\end{equation}
We show the agreement between this approximation and the critical embedding's mass in the next section (see figure \ref{fig:1q_m_star}).
%
\subsection{Numerical Results} \label{sec:numerics}
%
We now proceed to solve the equations of motion derived from the Lagrangian in equation \reef{eqt:legendre}.  To accomplish this, we use a shooting method.  For black hole embeddings, we choose a value for $\theta_0$ between 0 and $\pi/2$; this in turn determines the value of $\tilde{u}_\ast$ \emph{via} equation \reef{eqt:vanishing_locus} and the value of $\tilde{K}$ \emph{via} equation \reef{eqt:k1}.  The value of $\theta_0'$ is determined from the equation of motion itself.  For Minkowski embeddings, a value for $\tilde{u}_{\rm min}>1$ is chosen with $\theta(\tilde{u}_{\rm min}) = \pi/2$.   In order to avoid a conical singularity in the embedding~\cite{Karch:2006bv}, $\theta'(\tilde{u}_{\rm min}) = - \infty$.  These provide the necessary IR initial conditions for our shooting method, and we solve the equations of motion towards $\tilde{u} \to \infty$.  From the asymptotic behavior of $\theta$ described in equation \reef{eqt:asymptotic}, we extract the bare quark mass and the condensate.  We present some results of this scheme in figure \ref{fig:1q_c_vs_m} and in figure \ref{fig:1q_m_star}.  In particular, we find that the mass of the critical embedding provides a good estimate of the mass where the phase transition occurs.  Since calculating the mass of the critical embedding $\tilde{m}_\ast$ is simpler than finding the mass at the phase transition $\tilde{m}_{\rm cr}$, for the rest of this article, we simply use $\tilde{m}_\ast$ as an estimate of $\tilde{m}_{\rm cr}$.
\FIGURE{\epsfig{file=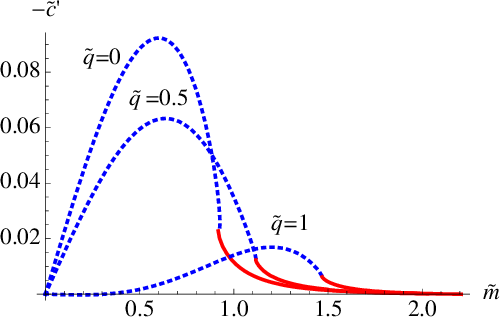,width=8cm}\epsfig{file=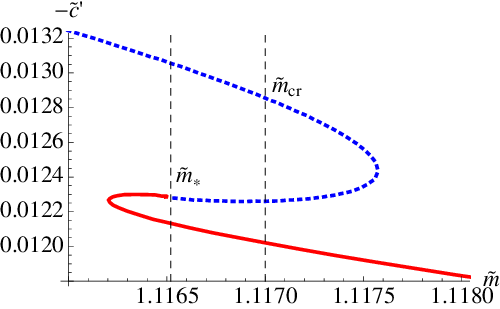,width=8cm}
\caption{\small The figure to the left shows the condensate as a function of the bare quark mass for three different $\tilde{q}$ values.  For the case $\tilde{q} = 1$, the condensate is negative for small (but finite) quark mass; this effect becomes more pronounce for higher $\tilde{q}$ values.  The figure to the right is a close up of the $\tilde{q}=0.5$ curve around the phase transition region.  $\tilde{m}_\ast$ denotes the mass at which the critical embedding occurs, and $\tilde{m}_{\rm cr}$ denotes the mass at which the phase transition occurs.  This is determined using an equal--area law~\cite{Albash:2006ew}.}
\label{fig:1q_c_vs_m}}
\FIGURE{\epsfig{file=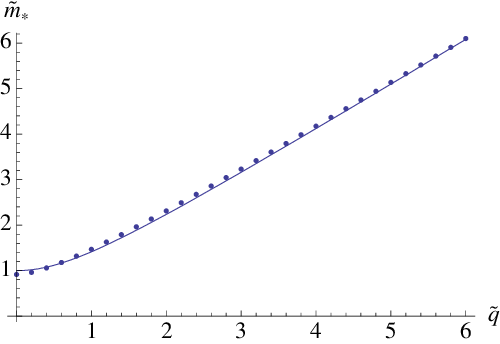,width=9cm}
 \caption{\small The dots correspond to the numerically calculated mass of the critical embedding, and the solid curve is a plot of equation \reef{eqt:m_star}.  We find a very good agreement between the two even for small $\tilde{q}$ where the analytical result is not supposed to be valid.}
 \label{fig:1q_m_star}}
%
\subsection{Quasinormal Mode Spectrum}\label{section:meson}
%
To study the quasinormal mode and meson spectrum, we need to consider quadratic (in $\alpha'$) fluctuations about our classical embeddings for the D7--brane.  To that end, we consider an expansion for the transverse fields of the form:
\begin{eqnarray}
\theta &=& \theta^{(0)} + 2 \pi \alpha'  \chi( \xi^a ) \ , \\
\phi_1 &=& \phi_1^{(0)} + 2 \pi \alpha' \Phi(\xi^a ) \ ,
\end{eqnarray}
where $\theta^{(0)}$ and $\phi_1^{(0)}$ are the classical embedding results found in the previous sections.  Expanding the action to second order in $2 \pi \alpha'$, we find the relevant terms for our fluctuation analysis are:
\begin{eqnarray}
- \mathcal{L}_{\chi^2} &=& \frac{1}{2} \sqrt{-g} g^{a b} G_{\theta \theta} \partial_a \chi \partial_b \chi -  \frac{1}{2} \sqrt{-g} g^{a b} \theta^{\prime 2} G_{\theta \theta}^2 g^{u u} \partial_a \chi \partial_b \chi \\
&& + \frac{1}{2} \chi^2 \left[ \partial_\theta^2 \sqrt{-g} - \partial_u \left(  \theta' G_{\theta \theta} g^{u u} \partial_\theta \sqrt{-g}  \right) \right] \ , \nonumber \\
- \mathcal{L}_{\Phi^2} &=& \frac{1}{2} \sqrt{-g} G_{\Phi \Phi} g^{a b} \partial_a \Phi \partial_b \Phi \ .
\end{eqnarray}
The indices $a, b$ range over the worldvolume coordinates, $g$ is the induced metric on the D7--brane at the classical level, and $G$ is the space--time metric.  For simplicity, we study the field $\Phi$ and change to a new coordinate $r = u^2$.  We take an ansatz of the form:
\begin{equation} \label{eqt:meson_ansatz}
\Phi = \Phi \left( \tilde{t} , \tilde{r} \right) = \phi \left( \tilde{r} \right) e^{i \tilde{\omega} \tilde{t}} \ ,
\end{equation}
where we have defined the dimensionless coordinates $\tilde{r} = r / u_H^2$, $ \tilde{t} = t u_H / R^2$, and $\tilde{\omega} = \omega R^2 / u_H$.  The equation of motion for $\phi \left( \tilde{r} \right)$ takes the form:
\begin{equation} \label{eqt:meson_eom}
\phi'' \left(\tilde{r} \right)+ \frac{1}{A} \left( i \tilde{\omega} B + A' \right) \phi' \left(\tilde{r} \right)+ \frac{1}{A} \left( C \tilde{\omega}^2 + \frac{i \tilde{\omega}}{2} B' \right) \phi\left(\tilde{r} \right) = 0 \ , 
\end{equation}
where $A$, $B$, and $C$ are given by:
\begin{displaymath}
A = \frac{1}{ \tilde{r}^2 \cos^6 \theta \sin \theta  g_{\theta}^{1/2}}  \frac{\tilde{g}_K^{3/2}}{\tilde{g}_r^{1/2}} \ , \quad B = \frac{\tilde{q} }{\tilde{f} \tilde{r}^2 \cos^6 \theta } \frac{\tilde{g}_K}{\tilde{g}_r}\ ,
\end{displaymath}
\begin{displaymath}
C = \frac{ \tilde{g}_\theta^{1/2} \left(- \tilde{q}^2 \tilde{K}^2 + \tilde{r}^2 \tilde{f}^2 \cos^6 \theta \left(\tilde{r} + \tilde{q}^2 \cos^2 \theta \right) \right) \sin \theta}{4 \tilde{f}^2 \tilde{r}^2 \cos^6 \theta} \frac{\tilde{g}_K^{1/2}}{ \tilde{g}_r^{3/2}} 
\end{displaymath}
\begin{displaymath}
\tilde{g}_K = - \tilde{K}^2 + \tilde{r}^2 \tilde{f} \cos^6 \theta \sin^2 \theta \ , \quad \tilde{g}_r = -1 +\tilde{q}^2 \tilde{r} \cos^2 \theta +  \tilde{r}^2 \ , \quad \tilde{g}_\theta = 1 + 4 \tilde{f} \theta'^2 \ , \quad \tilde{f} = -1 + \tilde{r} \tilde{q}^2 + \tilde{r}^2
\end{displaymath}
At this point, we choose to focus on the quasinormal modes by studying the fluctuations of the embeddings that cross the vanishing locus and fall into the event horizon.  Note that near $\tilde{r}_\ast = \tilde{u}_\ast^2$, we have:
\begin{displaymath}
A \Big|_{\tilde{r} \to \tilde{r}_\ast} \propto (\tilde{r} - \tilde{r}_\ast), \quad B \Big|_{\tilde{r} \to \tilde{r}_\ast} \propto (\tilde{r} - \tilde{r}_\ast)^0, \quad C \Big|_{\tilde{r} \to \tilde{r}_\ast} \propto (\tilde{r} - \tilde{r}_\ast)^{-1} \ .
\end{displaymath}
Using this, we find that the equation of motion, near $\tilde{r}_\ast$, reduces to:
\begin{equation}\label{eqt:meson1}
\phi'' \left(\tilde{r} \right) + \frac{D}{\tilde{r} - \tilde{r}_\ast} \phi' \left(\tilde{r} \right)+ \frac{E}{\left( \tilde{r} - \tilde{r}_\ast \right)^2} \phi\left(\tilde{r} \right)  = 0 \ .
\end{equation}
where $D$ and $E$ are constants depending on $\tilde{r}_\ast$.  We do not express these terms explicitly as they are particularly long and not illuminating.  Equation \reef{eqt:meson1} is particularly relevant since it is exactly of the form of ingoing/outgoing wave solutions~\cite{Hoyos:2006gb}.  The correct boundary condition at the vanishing locus is the ingoing solution.  A similar result was found in ref.~\cite{Albash:2007bq}.  However, there are indications that the gauge theory only sees the physics up to the vanishing locus. The constant $\tilde{K}$ is defined at $\tilde{u}_\ast$; the asymptotic behavior of the branes can be solved for by imposing initial conditions at $\tilde{u}_\ast$, so the bare quark mass and condensate can be identified without having to solve for the embedding up to the event horizon. From our results above, the same holds true for the meson spectrum. Therefore, one may be tempted to interpret the vanishing locus as an effective horizon that screens the physics of the D7--brane between the vanishing locus and the event horizon.  Of course, Ramond--Ramond (RR) charge conservation on the D7--brane requires us to continue the embedding up to the event horizon.  However, the phase diagram can be generated by applying boundary conditions at the vanishing locus.  Perhaps for the questions we are asking, the gauge theory is ignorant of the physics inside the vanishing locus, and hence the conical singularities we have encountered and the question of their resolution do not alter the phase diagrams we have presented.

One could worry whether the on--shell action of the black hole embeddings receives contributions from the region below the vanishing locus and therefore play an important role in determining the phase diagram.  However, in constructing the phase diagram, we employ an equal--area law, which follows from the fundamental definition of the free energy in terms of our thermodynamic variables, the bare quark mass and its thermodynamic conjugate the condensate. The on--shell action of the D7--brane does not fit with this prescription, suggesting a subtraction procedure in the presence of a vanishing locus that would be interesting to explore in future work.

A similar calculation can be performed for the field $\chi$, yielding the same properties for the vanishing locus.  We do a little more analytical work by considering the simple case of fluctuations around the $\theta = 0$ embedding.  In this case, the vanishing locus and the event horizon coincide.  The equation of motion takes the form:
\begin{equation} \label{eqt:meson_eom}
\phi''(\tilde{r} ) + \frac{ 3 \tilde{r}^2 + 2 \tilde{q}^2 r -1}{ \tilde{r} \tilde{f} } \phi' ( \tilde{r} ) + \frac{-3 + 3 \tilde{r}^2 + \tilde{r} \tilde{\omega}^2 + 4 \tilde{q}^2 \tilde{r} + \tilde{q}^2 \tilde{\omega}^2}{4 \tilde{f}^2 } \phi (\tilde{r}) = 0
\end{equation}
Near the event horizon, the equation reduces to:
\begin{equation}
\phi''(\tilde{r}) + \frac{1}{\tilde{r} - \tilde{r}_E} \phi'( \tilde{r}) + \frac{ \tilde{q}^6 + 4 \tilde{\omega}^2 + \tilde{q}^4 \left( \tilde{\omega}^2 - \sqrt{4+ \tilde{q}^4} \right)+ \tilde{q}^2 \left(4 +  \tilde{\omega}^2  \sqrt{4+ \tilde{q}^4} \right)}{8 \left(4 + \tilde{q}^4 \right)^{3/2} \left( \tilde{r} - \tilde{r}_E \right)^2} \phi (\tilde{r}) = 0
\end{equation}
This equation is solve by:
\begin{equation} \label{eqt:meson_bc}
\phi (\tilde{r} ) = c_1 \left( \tilde{r} - \tilde{r}_E \right)^{ i \tilde{\Omega}} + c_2  \left( \tilde{r} - \tilde{r}_E \right)^{- i \tilde{\Omega}}
\end{equation}
where $\tilde{\Omega} =  \frac{1}{2} \sqrt{ \frac{ - \tilde{q}^4 + \sqrt{4+ \tilde{q}^4} \tilde{\omega}^2 + \tilde{q}^2 \tilde{\omega}^2 + \tilde{q}^2 \sqrt{4 + \tilde{q}^4}}{ 2 \left( 4+ \tilde{q}^4 \right)}}$.  The first solution is the outgoing wave solution, and the second solution is the incoming wave solution.  When taking the $\tilde{q} \to 0$ limit, the solution reduces to the solution found for the (uncharged) AdS--Schwarzschild black hole~\cite{Hoyos:2006gb,Albash:2006ew}, which is a reassuring check of our result.  We proceed to solve for the spectrum numerically.  To accomplish this, we first point out that for large $\tilde{r}$, the equation of motion in equation \reef{eqt:meson_eom} asymptotes to:
\begin{equation}
\phi''( \tilde{r}) + \frac{3}{\tilde{r}} \phi'(\tilde{r}) + \frac{3}{4 \tilde{r}^2} \phi(\tilde{r}) = 0 \ .
\end{equation}
This has solutions given by:
\begin{equation}
\phi(\tilde{r}) = c_1 \tilde{r}^{-1/2} + c_2 \tilde{r}^{-3/2} \ .
\end{equation}
The first term corresponds to the non--normalizable term, and so that coefficient must be required to be zero.  Therefore, to proceed, we impose the incoming boundary condition in equation \reef{eqt:meson_bc} at the vanishing locus .  We scan through complex values of $\tilde{\omega}$ until the normalizable solution is found.  We present the solution for the lowest mode~\cite{Kruczenski:2003be} in figure \ref{fig:1q_meson}.\\
\FIGURE{\epsfig{file=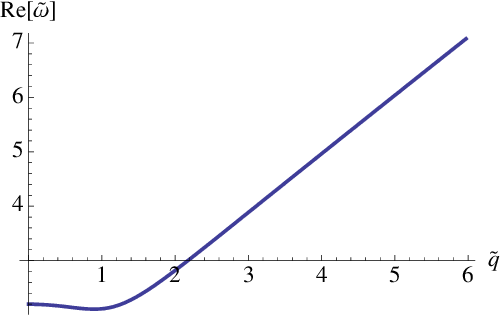,width=8cm}\epsfig{file=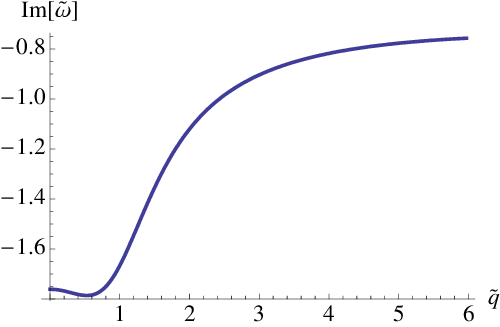,width=8cm}
\caption{\small The spectrum for the $\theta$ fluctuations about the $\theta=0$ embedding.  The real part of $\tilde{\omega}$ corresponds to the mass of the meson before it melts into the plasma, and the imaginary part of $\tilde{\omega}$ is proportional to the inverse of the lifetime of the meson~\cite{Hoyos:2006gb}.}
\label{fig:1q_meson}}
\indent In order to study the meson spectrum, we would need to solve equation \reef{eqt:meson_eom} about the Minkowski embeddings.  We emphasize that since the Minkowski embeddings close above the vanishing locus, normalizable solutions are possible, and the mesons associated with the fluctuations are stable.  Given the form of the equation, the numerical analysis is messy and slow.  However, we do not expect the physics for these fluctuations to be qualitatively different from what has already been seen before in similar cases \cite{Albash:2007bq}.  In particular, as the embeddings close further above the event horizon, the meson spectrum should approach the solution found for the pure AdS$_5 \times S^5$ background \cite{Kruczenski:2003be}. 
%
\subsection{Phase Diagram}
%
Using equations \reef{eqt:temperature} and \reef{eqt:mu_R}, we calculate the temperature and the chemical potential for the single charge:
\begin{equation}
T = \frac{u_H}{\pi R^2} \sqrt{1+\frac{\tilde{q}^4}{4}} \sqrt{\sqrt{1+\frac{\tilde{q}^4}{4}}-\frac{\tilde{q}^2}{2}} \ , \quad \mu_R = \frac{u_H}{R^2} \tilde{q} \left( \sqrt{1+ \frac{\tilde{q}^4}{4}}  - \frac{\tilde{q}^2}{2} \right) \ .
\end{equation}
In particular, we find that these two thermal quantities are not entirely independent of each other.  To clarify this point, we consider the ratio of these two terms:
\begin{equation}
\frac{T}{\mu_R} = \frac{\sqrt{1+ \frac{\tilde{q}^4}{4}}}{ \pi \tilde{q} \sqrt{\sqrt{1+\frac{\tilde{q}^4}{4}}-\frac{\tilde{q}^2}{2}}} \geq \frac{\sqrt{2}}{\pi} \ ,
\end{equation}
where the inequality is saturated at $\tilde{q}^2 = 2/ \sqrt{3}$.  The fact that the ratio does not go to zero is not surprising, since the extremal black hole occurs for $u_H = 0$ and not for any particular value of $\tilde{q}$.  Therefore, in the $\left(\mu_R, T \right)$ plane, only states above the line $T = \mu_R \sqrt{2} / \pi$ are physical.  In addition, the cases of $\tilde{q} \leq 2/ \sqrt{3}$ and $\tilde{q} \geq 2/ \sqrt{3}$ cover the same values of the ratio; therefore, one needs only one interval to cover all physical states.  We choose to consider only values of $\tilde{q}$ satisfying $\tilde{q}^2 \leq 2/ \sqrt{3}$.
To proceed with drawing the phase diagram, we derive the relevant dimensionless quantities.  Using that $R^4 = 4 \pi g_s \alpha^{\prime 2} N_c= \lambda \alpha^{\prime 2}$ and $m_q = \tilde{m} u_H / 2 \pi \alpha'$, we write as dimensionless physical parameters:
\begin{eqnarray}
\sqrt{\lambda} \frac{\mu_R}{m_q} &=& \frac{2 \pi \tilde{q}}{\tilde{m}} \left(\sqrt{1+ \frac{\tilde{q}^4}{4}} - \frac{\tilde{q}^2}{2} \right) \ , \\
\sqrt{\lambda} \frac{T}{m_q} &=& \frac{2}{\tilde{m}} \sqrt{1+ \frac{\tilde{q}^4}{4}} \sqrt{\sqrt{1+ \frac{\tilde{q}^4}{4}} - \frac{\tilde{q}^2}{2}} \ .
\end{eqnarray}
With these definitions in mind, we present the phase diagram of the single charge case in figure \ref{fig:1q_phase_diagram}.
\FIGURE{\epsfig{file=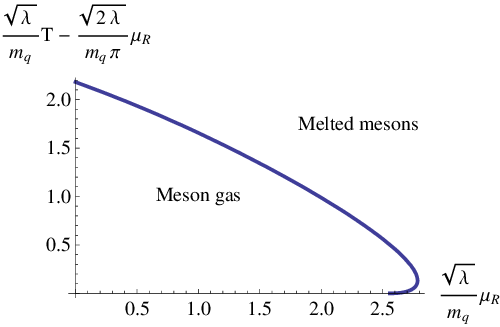,width=10cm}
 \caption{\small The phase diagram for the single charge case.}
 \label{fig:1q_phase_diagram}}
 %
\section{Three Equal Charges}
\subsection{Properties of the embeddings}
\label{sec: three charges}
%
When we set all three charges to be equal~\cite{Chamblin:1999tk}, ({\it i.e.,} $q_1=q_2=q_3=q$ in equation~\reef{eqt: decoupled spinning brane metric}), the metric is given by:
\begin{eqnarray} \label{eqt: three charge metric}
ds^2/\alpha' &=&
-\mathcal{H}^{-2} f dt^2 + \mathcal{H}f^{-1} du^2 + \mathcal{H}\frac{u^2}{R^2}d\vec{x}\cdot d\vec{x} \nonumber\\
&&+\sum_{i=1}^3\left\{R^2d\mu_i^2 +\mu_i^2\left(R d\phi'_i -A_t' dt\right)^2\right\} \ ,
\end{eqnarray}
where
\begin{equation}
\mathcal{H} = 1+ \frac{q^2}{u^2} \ , \quad 
f =  \frac{\mathcal{H}^3 u^2}{R^2} - \frac{u_H^4}{u^2 R^2} \ , \quad 
A_t = \frac{u_H^2 q}{ R(u^2+  q^2)} -  \frac{u_H^2 q}{ R(u_E^2+  q^2)} \ .
\end{equation}
The event horizon $u_E$ satisfies the equation:
\begin{equation}
\left(u_E+q^2 \right)^3 - u_H^4 u_E^2 = 0
\end{equation}
We proceed in much the same way as in the single charge case, so we will be brief.  We use an ansatz as in equation \reef{eqt:ansatz}, and the D7--brane probe action is given by a similar object as \reef{eqt:action} but with:
\begin{equation}
\tilde{\mathcal{L}} =\tilde{u}^3 \mathcal{H} \cos^3 \theta(\tilde{u}) \sqrt{  \left(1- \frac{\tilde{q}^2}{\tilde{f}(\tilde{u})} \sin^2 \theta(\tilde{u}) \right) \left(1+ \frac{\tilde{f}(\tilde{u})}{\tilde{u}^4 \mathcal{H}}\theta'(\tilde{u})^2 \right) + \frac{\tilde{f}(\tilde{u})}{\tilde{u}^4 \mathcal{H}} \sin^2 \theta(\tilde{u}) \phi_1'(\tilde{u})^2} \ , 
\end{equation}
and $\tilde{f}(\tilde{u}) = \left(\tilde{u}+\tilde{q}^2 \right)^3 - \tilde{u}^2$.  The equation of motion for $\phi_1$ can be integrated to give a constant of motion:
\begin{equation}
\frac{ \tilde{f}(\tilde{u}) \cos^3 \theta(\tilde{u}) \sin^2 \theta(\tilde{u}) \phi_1'(\tilde{u})}{\tilde{u} \sqrt{\left(1 -\frac{\tilde{q}^2}{\tilde{f}(\tilde{u})} \sin^2 \theta(\tilde{u}) \right) \left(1 + \frac{\tilde{f}(\tilde{u})}{ \mathcal{H} \tilde{u}^4} \theta'(\tilde{u})^2 \right) + \frac{\tilde{f}(\tilde{u})}{\tilde{u}^4 \mathcal{H}} \sin^2 \theta(\tilde{u}) \phi_1'(\tilde{u})^2}} = - \tilde{K} \ .
\end{equation}
We again choose to work with the parameter $\tilde{K}$ instead of the field $\phi_1$ by performing a Legendre transformation:
\begin{equation}
\tilde{I} = \frac{1}{\sin \theta(\tilde{u}) \tilde{f}(\tilde{u})} \sqrt{ \left(\tilde{f}(\tilde{u}) - \tilde{q}^2 \sin^2 \theta(\tilde{u}) \right) \left( -\tilde{K}^2 + \tilde{f}(\tilde{u}) \mathcal{H} \tilde{u}^2 \cos^6 \theta (\tilde{u}) \sin^2 \theta(\tilde{u}) \right) \left( \mathcal{H} \tilde{u}^4 + \tilde{f}(\tilde{u}) \theta'(\tilde{u})^2 \right)}
\end{equation}
The vanishing locus, denoted by $\tilde{u}_\ast$ satisfies:
\begin{equation}
\left( \tilde{u}_\ast^2 + \tilde{q}^2 \right)^3 - \tilde{u}_\ast^2 - \tilde{q}^2 \sin^2 \theta_0 = 0 \ ,
\end{equation}
where $\theta_0 \equiv \theta(\tilde{u}_\ast)$.  The requirement that the action remain real beyond the vanishing locus fixes the value of $\tilde{K}$ to be:
\begin{equation} \label{eqt:k2}
\tilde{K}^2 =  \tilde{f}(\tilde{u}_\ast) \left(\tilde{q}^2 + \tilde{u}^2_\ast \right) \cos^6 \theta_0 \sin^2 \theta_0 \ .
\end{equation}
Following the procedure in section \ref{sec:numerics}, we can solve the equation of motion for $\theta(\tilde{u})$ using a shooting method.  We find the same two types of embeddings, black hole and Minkowski embeddings, and the black hole embeddings once again exhibit a conical singularity above a certain $\theta_0$.  We present some results of the condensate in figure \ref{fig:3q_c_vs_m}.  
\FIGURE{\epsfig{file=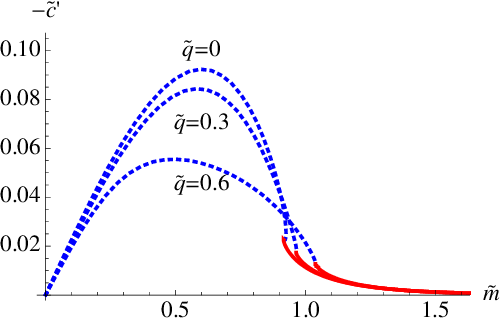,width=8cm}\epsfig{file=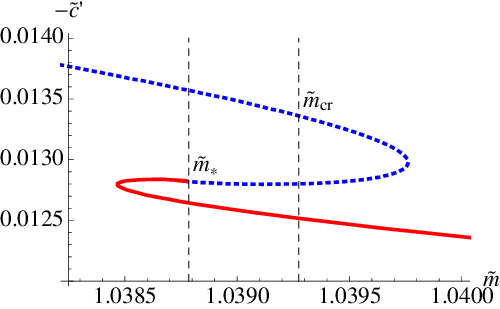,width=8cm}
\caption{\small The figure to the left shows the condensate as a function of the bare quark mass for three different $\tilde{q}$ values.  The figure to the right shows a close up of the phase transition region for the case of $\tilde{q} = 0.6$.}
\label{fig:3q_c_vs_m}}
%
\subsection{Quasinormal Mode Spectrum}
In order to study the quasinormal mode and meson spectrum, we follow the same procedure described in section \ref{section:meson}.  We take the same ansatz as in equation \reef{eqt:meson_ansatz} for the field $\Phi$ corresponding to fluctuations in the $\phi_1$ direction, but in this case the variable $\tilde{r}$ is given by $\tilde{r} = \tilde{u}^2 + \tilde{q}^2$.  The equation of motion for $\phi(\tilde{r})$ has the exact form as in equation \reef{eqt:meson_eom}, but now we have:
\begin{displaymath}
A = \frac{1}{\sin \theta \cos^6 \theta r \tilde{g}_\theta^{1/2}} \frac{\tilde{g}_K^{3/2}}{\tilde{g}_r^{1/2}} \ , \quad B = \frac{\tilde{q} \tilde{K}}{\cos^6 \theta \tilde{f} } \frac{\tilde{g}_K}{\tilde{g}_R} \ , \quad C = \frac{ r \left(- \tilde{q}^2 \tilde{K}^2 + \tilde{r} \tilde{f}^3 \cos^6 \theta \right) \sin \theta \tilde{g}_\theta^{1/2}}{4 \cos^6 \theta \tilde{f}^2} \frac{ \tilde{g}_K^{1/2}}{\tilde{g}_r^{3/2}} \ ,
\end{displaymath}
\begin{displaymath}
\tilde{g}_K = - \tilde{K}^2 + \tilde{r} \tilde{f} \cos^6 \theta \sin^2 \theta \ , \quad \tilde{g}_r = \tilde{r}^3 - \tilde{r} + q^2 \cos^2 \theta \ , \quad \tilde{g}_\theta = r + 4 \tilde{f} \theta'^2 \ , \tilde{f} = \tilde{r}^3 - \tilde{r} + \tilde{q}^2 \ .
\end{displaymath}
We once again have that near $\tilde{r}_\ast = \tilde{u}_\ast^2 + \tilde{q}^2$, we have:
\begin{displaymath}
A \Big|_{\tilde{r} \to \tilde{r}_\ast} \propto (\tilde{r} - \tilde{r}_\ast), \quad B \Big|_{\tilde{r} \to \tilde{r}_\ast} \propto (\tilde{r} - \tilde{r}_\ast)^0, \quad C \Big|_{\tilde{r} \to \tilde{r}_\ast} \propto (\tilde{r} - \tilde{r}_\ast)^{-1} \ .
\end{displaymath}
Using this, we find that the equation of motion, near $\tilde{r}_\ast$, reduces to:
\begin{equation}\label{eqt:meson}
\phi'' \left(\tilde{r} \right) + \frac{D}{\tilde{r} - \tilde{r}_\ast} \phi' \left(\tilde{r} \right)+ \frac{E}{\left( \tilde{r} - \tilde{r}_\ast \right)^2} \phi\left(\tilde{r} \right)  = 0 \ .
\end{equation}
Therefore, the arguments from section \ref{section:meson} for the single charge case hold for the three charge case as well.
%
\subsection{Phase Diagram}
%
To proceed with drawing the phase diagram, we calculate the temperature and the chemical potential for the background:
\begin{equation}
T = \frac{u_H \tilde{u}_E}{2 \pi R^2} \left(2 \tilde{u}_E^2 + 3 \tilde{q}^2 - \frac{\tilde{q}^6}{\tilde{u}_E^4} \right) \ , \quad \mu_R = \frac{u_H}{R^2} \frac{\tilde{q}}{\tilde{u}_E^2 + \tilde{q}^2} \ .
\end{equation}
What is different about this case is that we can achieve zero temperature without sending $u_H \to 0$.  The extremal black hole is achieved when $\tilde{f}(\tilde{u})$ has a double root, which occurs at a value of:
\begin{equation}
\tilde{q}_{\rm ex}^2 = \frac{2 \sqrt{3}}{9}  \quad \leftrightarrow \quad \tilde{u}_E^2 = \frac{\sqrt{3}}{9}\ .
\end{equation}
With this in mind, the chemical potential at zero temperature is simply given by:
\begin{equation}
\mu_R (T = 0) = \frac{u_H}{R^2} \sqrt{\frac{2 \sqrt{3}}{3}} \ .
\end{equation}
Therefore, for the three charge case at zero temperature, $u_H$ purely controls the chemical potential.  We define dimensionless physical parameters:
\begin{eqnarray}
\sqrt{\lambda} \frac{\mu_R}{m_q} &=& \frac{2 \pi}{\tilde{m}} \frac{\tilde{q}}{\tilde{u}_E^2 + \tilde{q}^2} \ , \\
\sqrt{\lambda} \frac{T}{m_q} &=& \frac{\tilde{u}_E^2}{\tilde{m}} \left(2 \tilde{u}_E^2 + 3 \tilde{q}^2 - \frac{\tilde{q}^6}{\tilde{u}_E^4} \right)
\end{eqnarray}
Note that it is possible to find $\tilde{u}_E$:
\begin{equation}
\tilde{u}_E^2 = -\tilde{q}^2 + \frac{2 \ 3^{1/3} + 2^{1/3} \left(-9 \tilde{q}^2 + \sqrt{ 81 \tilde{q}^4 -12 } \right)^{2/3}}{6^{2/3} \left(-9 \tilde{q}^2 + \sqrt{ 81 \tilde{q}^4 -12} \right)^{1/3}} \ .
\end{equation}
With this in mind, we present the phase diagram for the three charge case in figure \ref{fig:3q_phase}.
\FIGURE{\epsfig{file=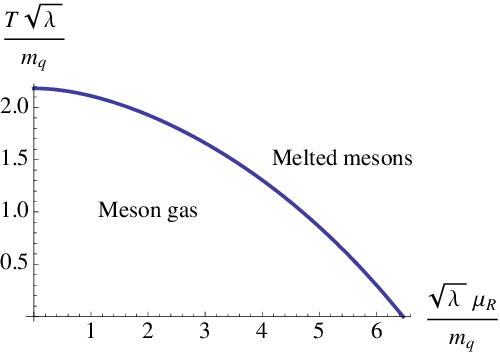,width=9cm}
\caption{\small The phase diagram for the three equal charge case.}
\label{fig:3q_phase}}
\subsection{The Physical Meaning of $\tilde{K}$}
We now turn to interpreting the physical meaning of the constant $\tilde{K}$, expressed in equations \reef{eqt:k1} and \reef{eqt:k2} for the single charge and three charge cases respectively.  We write the Hamiltonian of the system:
\begin{equation}
\mathcal{H} = \frac{1}{2} \left( m_q \mathcal{O}_q + m_q^\ast \mathcal{O}_q^\ast \right) + \mathcal{H}_0 \ ,
\end{equation}
where $\mathcal{H}_0$ includes all terms that do not depend on the bare quark mass.  If $m_q$ is real, then $\langle \mathcal{O}_q \rangle$ corresponds to the condensate of the theory.  If $m_q$ is not real, we write:
\begin{equation}
m_q = | m_q | e^{i \phi_m} \ , \quad \langle \mathcal{O}_q \rangle = |\langle \mathcal{O}_q \rangle| e^{- i \phi_c } \ .
\end{equation}
If this is the case, then the condensate of the theory is given by:
\begin{equation}
\frac{1}{2} \left\langle m_q \mathcal{O}_q + m_q^\ast \mathcal{O}_q^\ast \right\rangle = | m_q | |  \langle \mathcal{O}_q \rangle | \cos \Delta \phi \ ,
\end{equation}
where $\Delta \phi = \phi_m - \phi_c$.  Now we identify the expectation value of the Hamiltonian with the regularized action \emph{via}:
\begin{equation}
\langle \mathcal{H} \rangle = - S_{\rm reg} \left[ \tilde{L}, \phi_1 \right]
\end{equation}
where $\tilde{L}$ is the asymptotic separation of the D3 and D7--branes and $\phi_1$ is the angle in the transverse plane to the D7--brane.  The asymptotic separation is in turn related to the bare quark mass \emph{via} $\tilde{L}_\infty = \tilde{m}$ and equation \reef{eqt:m}.  Using the standard variational argument, we relate the condensate to $\tilde{L}_\infty$:
\begin{equation}
|  \langle \mathcal{O}_q \rangle | \cos \Delta \phi = - 2 \pi \alpha' \frac{\delta S_{\rm reg}}{\delta \partial_u L} \Big|_{\infty} = 2 \pi \alpha' u_H^3 2 \pi^2 N_f T_{\rm D7}  V \left(- 2 \tilde{c}' \right)\ .
\end{equation}
In a T--dual description, the separation between the D3 and D7--branes corresponds to the expectation value of the $U(1)$ gauge field on the D7--brane~\cite{Polchinski:1998rq}.  Rotations in the transverse planes, parametrized by $\phi_1$, rotate components of the D7--brane's gauge field.  In turn, rotations in the transverse plane correspond to rotating the components of the bare quark mass.  Therefore, we write:
\begin{equation}
 | m_q | |  \langle \mathcal{O}_q \rangle | \sin \Delta \phi = \frac{\delta S_{\rm reg}}{\delta \phi_1'} \Big|_{\infty} = u_H^4 2 \pi^2 N_f T_{\rm D7} V \tilde{m}^2 \frac{\tilde{K}}{2} \ .
\end{equation}
This in turn means that:
\begin{equation}
\tan \Delta \phi =- \frac{\tilde{m} \tilde{K}}{4 \tilde{c}'}
\end{equation}
We see that the constant $\tilde{K}$ is directly related to the phase difference between the bare quark mass and the condensate of the dual gauge theory.  Also, since $\Delta \phi = 0$ for Minkowski embeddings and $\Delta \phi \neq 0$ for black hole embeddings, $\Delta \phi$ is a natural order parameter for the first order phase transition that we study.  This situation is very similar to the case of having an external electric field, where the order parameter of the phase transition is a global electric current~\cite{Karch:2007pd,Albash:2007bq,Filev:2008xt}.
%
\appendix
\section{The Condensate} \label{appendix:condensate}
%
We present arguments to justify the additional term proportional to $\tilde{m} \tilde{q}^2$ in equation \reef{eqt:condensate} for $\tilde{c}'$.  We first present the geometrical argument.   In order to be able to identify the condensate, we want to restore the symmetry of the $S^{5}$ as much as possible as we move away from the horizon.  By looking at the distorted $S^{5}$ part in equation~\reef{eqt: decoupled spinning brane metric}, one can see that, as we start moving away from the horizon (i.e. when we see the effect of $u_H$ less and less), it starts conformally looking like:
\begin{equation}
{ds_{S_5}}^2=\sum_{i=1}^3{{\mathcal{H}_i} \left(d\mu_i^2+d\phi_i^2\mu_i^2\right)} \ ,
\end{equation}
where
\begin{eqnarray}
\mu_1 &=& \sin\theta \ , \quad 
\mu_2 = \cos\theta \sin \psi \ , \quad 
\mu_3 = \cos\theta \cos \psi \ .
\end{eqnarray}
Following ref.~\cite{Russo:1998mm}, the change of variables that would restore the symmetry is given by:
\begin{eqnarray}
y_{1}&=&\sqrt{r^2+q_{1}^2} \ \mu_1\cos\phi_1\ ,\quad 
y_{2}=\sqrt{r^2+q_{1}^2} \ \mu_1\sin\phi_1 \ , \quad
y_{3}=\sqrt{r^2+q_{2}^2} \ \mu_2\cos\phi_2\ ,\quad \nonumber \\
y_{4}&=&\sqrt{r^2+q_{2}^2} \ \mu_2\sin\phi_2 \ ,\quad
y_{5}=\sqrt{r^2+q_{3}^2} \ \mu_3\cos\phi_3 \ ,\quad
y_{6}=\sqrt{r^2+q_{3}^2} \ \mu_3\sin\phi_3 \ .
\end{eqnarray}
In view of this, for the single charge case, one defines the following radial coordinate:
\begin{equation}
r=\frac{u^2+\sqrt{u^2(u^2+q^2)-u_{H}^4}}{\sqrt{q^2+2u^2+2\sqrt{u^2(u^2+q^2)-u_{H}^4}}} \ .
\end{equation}
This allows us to rewrite:
\begin{equation}
f^{-1}du^2+R^2d\theta^2=R^2\left(\frac{dr^2}{r^2+q^2}+d\theta^2\right) \ .
\end{equation}
We can now define new coordinates:
\begin{eqnarray}
\widetilde{L}=\sqrt{r^2+q^2}\sin\theta\ ,\quad
 \rho=r\cos\theta\nonumber \ ,
 \end{eqnarray}
so that we have:
 \begin{equation}
R^2(\frac{dr^2}{r^2+q^2}+d\theta^2)=\frac{R^2}{\sqrt{4q^2\rho^2+(L^2+\rho^2-q^2)^2}}(d\widetilde{L}^2+d\rho^2) \ .
\end{equation}
Then from the asymptotic behavior of $r$ for large $u$:
\begin{equation}
\sqrt{r^2+q^2}=u+\frac{q^2}{2u}+\dots \ ,
\end{equation}
we can argue that $\widetilde{L}$ has the expansion:
\begin{equation}
\widetilde{L}=m+\frac{c+m\frac{q^2}{2}}{\rho^2}+\dots \ .
\end{equation}
Using the standard argument from ref.~\cite{Kruczenski:2003uq} we
can show that:
\begin{equation}
\langle\bar{\Psi}\Psi\rangle\sim-c-m\frac{q^2}{2} \ .
\end{equation}
So, in these coordinates that restore the symmetry of the $S^{5}$, the na\"ive condensate value gets a constant subtraction.  A similar argument can be made for the three charge case.  To show that, we introduce a new radial coordinate $r(u)$ satisfying:
\begin{equation}
\mathcal{H} f^{-1}du^2+R^2=\frac{R^2}{r^2}(dr^2+r^2d\theta^2) \label{eqn2} \ ,
\end{equation}
and the corresponding radial coordinates are:
\begin{eqnarray}
\widetilde{L}=r\sin{\theta}\ ,\quad
\widetilde{\rho}=r\cos{\theta} \ .
\end{eqnarray}
Using that $r(u)$ from equation~\reef{eqn2} has the asymptotic behavior:
\begin{equation}
r(u)=u+\frac{q^2}{2u}+O\left(\frac{1}{u^3}\right) \ ,
\end{equation}
one can show that:
\begin{equation}
\widetilde{L}=m+\left(c+m\frac{q^2}{2}\right)\frac{1}{\rho^2}+O\left(\frac{1}{\rho^4}\right) \ .
\end{equation}
Note also that one can write:
\begin{equation}
 \mathcal{H} f^{-1}du^2+R^2=\frac{R^2}{L^2+\rho^2}(d\rho^2+dL^2) \ ,
\end{equation}
and following the standard argument presented in ref.~\cite{Kruczenski:2003uq} one can show that:
\begin{equation}
\langle \bar{\Psi}\Psi\rangle\sim-c-m\frac{q^2}{2} \ .
\end{equation}

To derive the correct normalization constants, we use the holographic dictionary.  The condensate can be derived \emph{via}:
\begin{equation}
\langle \bar{\Psi} \Psi \rangle = \frac{\delta \mathcal{F}}{ \delta m_q} \ ,
\end{equation}
where $\mathcal{F}$ is the free energy of the system.  Following ref.~\cite{Karch:2005ms}, we write:
\begin{equation}
\frac{\delta \mathcal{F}}{ \delta m_q} = 2 \pi \alpha' \lim_{u_{\rm max} \to \infty} \frac{u^3}{\sqrt{-\gamma}} \frac{\delta { \left(- S_{\rm D7 }^{\rm reg} \right)}}{\delta \theta(u_{\rm max})} \ .
\end{equation}
In this case, $\sqrt{-\gamma}$ is the determinant of the boundary metric, given by:
\begin{equation}
\sqrt{-\gamma} = u_{\rm max}^4 + 2 q^2 u_{\rm max}^2 + \frac{1}{2} \left( 2 q^4 - 1 \right) \ .
\end{equation}
The regularized action $S_{\rm D7}^{\rm reg}$ includes the necessary counterterms to regularize the on--shell action~\cite{Karch:2005ms}:
\begin{equation}
-S_{\rm D7}^{\rm reg} = -S_{\rm D7}\Big|_{\rm on-shell} + 2 \pi^2 N_f T_{\rm D7} \int d^4 x \sqrt{-\gamma} \left(- \frac{1}{4} + \frac{1}{2} \theta(u_{\rm max})^2 - \frac{5}{12} \theta(u_{\rm max})^4 \right) \ .
\end{equation}
Using that the variation of the on--shell action is given by a boundary term and using that $\theta(u_{\rm max}) = m / u_{\rm max} + c / u_{\rm max}^3$, we arrive to the result that, for both the single and three equal charge case,:
\begin{equation}
\langle \bar{\Psi} \Psi \rangle = - 8 \pi^3 \alpha' V N_f T_{\rm D7} u_H^3 \left(\tilde{c} + \tilde{m} \frac{ \tilde{q}^2}{2} \right)  \ .
\end{equation}
\acknowledgments
We are grateful to Nick Evans, Rob Myers, and Andreas Karch for
comments. This research is supported by the US Department of Energy.

\bibliographystyle{JHEP}
\providecommand{\href}[2]{#2}\begingroup\raggedright\endgroup

\end{document}